\documentstyle[12pt,epsfig]{article}

\newcounter{subequation}[equation]

\makeatletter
\expandafter\let\expandafter\reset@font\csname reset@font\endcsname
\newenvironment{subeqnarray}
  {\arraycolsep1pt
    \def\@eqnnum\stepcounter##1{\stepcounter{subequation}{\reset@font\rm
      (\theequation\alph{subequation})}}\eqnarray}%
  {\endeqnarray\stepcounter{equation}}
\makeatother

\newcounter{statement}
\newenvironment{statement}[4]
  {\par\refstepcounter{statement}
    \noindent#1#2 \arabic{statement} #4\unskip: #3}{\par\vspace{2mm}}

\newenvironment{statement*}[4]
  {\par\noindent#1#2 #4\unskip: #3}{\par\vspace{2mm}}

\newcommand{\labelcaption}[2]{\caption[#1]{\label{#1}#2}}

\newcommand{\bu}{{\bar{U}}}
\newcommand{\bb}{{\bar{B}}}

\begin{document}

\hbox to\hsize{%

  \vbox{\hbox{Submitted to}\hbox{\sl Nucl.\ Phys.\ B}}\hfil
  \vbox{%
        \hbox{MPI-PhT/97-20}%
        \hbox{BUTP-97/08}%
        \hbox{gr-qc/9703047}%
        \hbox{March 18, 1997}%
       }}
\vspace{5mm}
\begin{center}
\LARGE\bf
Mass inflation and chaotic behaviour inside hairy black holes

\vskip5mm
\large Peter Breitenlohner$~{^\dagger}$,
George Lavrelashvili$~{^\ddagger}$
\footnote{On leave of absence from Tbilisi
Mathematical Institute, 380093 Tbilisi, Georgia}\\
{\normalsize and} Dieter Maison$~{^\dagger}$

\vspace{3mm}
{\small\sl
Max-Planck-Institut f\"ur Physik$~{^\dagger}$\\
--- Werner Heisenberg Institut ---\\
F\"ohringer Ring 6\\
80805 Munich (Fed.\ Rep.\ Germany)

\vspace{3mm}
Institute for Theoretical Physics$~{^\ddagger}$\\
University of Bern\\
Sidlerstrasse 5 \\
CH-3012 Bern, Switzerland\\}

\end{center}
\begingroup \addtolength{\leftskip}{1cm} \addtolength{\rightskip}{1cm}

\begin{center}\bf Abstract\end{center}
\vspace{1mm}\noindent
We analyze the interior geometry of static, spherically symmetric black
holes of the Einstein-Yang-Mills-Higgs theory.  Generically the
solutions exhibit a behaviour that may be described as ``mass
inflation'', although with a remarkable difference between the cases
with and without a Higgs field.  Without Higgs field the YM field
induces a kind of cyclic behaviour leading to repeated cycles of mass
inflation -- taking the form of violent explosions -- interrupted by
quiescent periods and subsequent approaches to an almost Cauchy horizon.
With the Higgs field no such cycles occur.  In addition there are
non-generic families with a Schwarzschild resp.\ Reissner-Nordstr{\o}m
type singularity at $r=0$.

\endgroup
\vspace{1cm}

\section{Introduction}\label{chint}

Most of the knowledge on the interior geometry of static black holes
derives from the well-known exact solutions like the Schwarzschild (S)
and Reissner-Nordstr{\o}m (RN) solution.  Whereas for the S-solution the
radial coordinate stays time-like inside the horizon all the way down to
the central singularity, the RN-solution exhibits a second -- inner --
horizon, behind which the radial coordinate becomes space-like again.
This inner horizon is a so-called Cauchy horizon, beyond which the
evolution of any matter system is influenced by data causally
disconnected from the earlier history \cite{HE}.  Furthermore, while the
outer horizon is an infinite redshift surface, the Cauchy horizon is a
surface of infinite blueshift.  This suggests, that the Cauchy horizon
should be unstable against small perturbations and a space-like
singularity should form replacing the former Cauchy horizon \cite{IP}.
In fact, perturbative studies indicate an exponential growth of the
mass-function close to the inner horizon -- a phenomenon dubbed ``mass
inflation'' \cite{IP}.  However, as yet no unanimous conclusion about
the reliability and genericity of these perturbative results seems to
have been obtained \cite{BI}.  In this situation it is definitely of
interest to investigate the internal structure of black holes with other
forms of matter like Yang-Mills and Higgs fields.  This is the
subject of the present study.  Compared to the previously mentioned
perturbative analysis our work is ``exact'', although at least partially
based on numerical results.

One may summarize these results saying that generically no Cauchy
horizon forms, because its existence requires fine-tuning of the initial
data at the outer horizon.  Solutions with a Cauchy horizon are in fact
obtained through such fine-tuning, leading to a RN-type singularity at
$r=0$.  Generically, however, the solutions exhibit ``mass-inflation'',
although with a remarkable difference between the cases with and without
a Higgs field.  The behaviour is actually much simpler with the
inclusion of a Higgs field.  In this case the mass function shows
generically exponential growth all the way down to $r=0$.  On the other
hand in the seemingly simpler case without the Higgs field the situation
is more involved.  The YM field induces a kind of cyclic behaviour
leading to repeated cycles of mass inflation -- taking the form of
violent ``explosions'' -- interrupted by quiescent periods and
subsequent approaches to an almost Cauchy horizon.  This behaviour is
particularly spectacular due to the fantastic growth of the mass function
during these explosions by hundreds of orders of magnitude, a phenomenon
unprecedented in standard physical problems.  Actually, these explosions
become exponentially more violent after each cycle such that it is
practically impossible to follow more than one or two of them
numerically.

A further novel phenomenon is a kind of chaotic behaviour generated by a
hierarchical structure of families of non-abelian RN-type (NARN) and
non-abelian S-type (NAS) solutions separating generic solutions with cyclic
(EYM) or acyclic (EYMH) behaviour.

Our numerical results exhibiting these claims are predominately based on
the EYM system, but we have no doubt that we could easily establish
similarly convincing evidence in the EYMH case.

In this discussion the global behaviour outside the horizon was ignored.
Clearly it is of interest to find out what is the behaviour of
asymptotically flat solutions, which for the EYM theory constitute a
discrete set of 1-parameter families -- described by curves in the
$(W_h,r_h)$ plane \cite{Bizon,BFM}.  Obviously without further restriction
they will show the generic behaviour inside the horizon, although for
certain discrete points S-type resp.\ RN-type singularities at $r=0$ are
possible.

In a recent paper \cite{DGZ} Donets et al.\ presented their results on
the internal structure of static, spherically symmetric black holes of
the Einstein-Yang-Mills (EYM) theory (without Higgs field).  Our results
for this particular case essentially agree with the findings of Donets
et al., although we differ in some details.  In particular the chaotic
stucture we find close to the basic NARN solutions was not observed in
\cite{DGZ}.

\section{Ansatz and Field Equations}\label{chapans}

The contents of this chapter are essentially (up to minor changes in
notation) a copy from our earlier paper \cite{BFM2}, which we include
for the convenience of the reader.

For the static, spherically symmetric metric we use the parametrization
\begin{equation}\label{Metr}
ds^2=A^2 Bdt^2-{dR^2\over B} -r^2(R)d\Omega^2\;,
\end{equation}
with $d\Omega^2=d\theta^2+\sin^2\theta d\varphi^2$
and three independent functions $A$, $B$, $r$ of a radial coordinate
$R$, which has, in contrast to $r$, no geometrical significance.
It is common to express $B$ through the ``mass function'' $m$ defined by
$B=1-2m/r$.

For the {\sl SU(2)\/} Yang-Mills field $W_\mu^a$ we use the
standard minimal spherically symmetric (purely `magnetic') ansatz
\begin{equation}\label{Ans}
W_\mu^a T_a dx^\mu=
  W(R) (T_1 d\theta+T_2\sin\theta d\varphi) + T_3 \cos\theta
d\varphi\;,
\end{equation}
and for the Higgs field we assume the form
\begin{equation}\label{AnsH}
\Phi^aT_a=H(R)n^aT_a\;,
\end{equation}
where $T_a$ denote the generators of ${\sl SU(2)\/}$ in the adjoint
representation.
Plugging these ans\"atze into the EYMH action results in
\begin{equation}\label{Action}
S=-\int dR A
  \Bigl[
  {1\over2}\Bigl(1+B((r')^2
   +{(A^2 B)'\over 2A^2 B}(r^2)')\Bigr)- Br^2V_1-V_2
\Bigr]\;,
\end{equation}
with
\begin{equation}
V_1={(W')^2\over r^2}+{1\over2}(H')^2\;,
\end{equation}
and
\begin{equation}
V_2={(1-W^2)^2\over2r^2}+
{\beta^2r^2\over8}(H^2-\alpha^2)^2+W^2H^2\;.
\end{equation}
Through a suitable rescaling we have achieved that the action depends
only on
the dimensionless parameters $\alpha$ and $\beta$ representing the
mass ratios $\alpha=M_W\sqrt G/g=M_W/g
M_{\rm Pl}$
and $\beta=M_H/M_W$ ($g$ denoting the gauge coupling and $G$ Newtons
constant).\footnote{
The usual quantum definition of
$M_{\rm Pl}\equiv\sqrt{\hbar c\over G}$ is obtained from our classical
expression replacing the dimensionful classical gauge coupling $g^2$
by the dimensionless  ${g^2\over\hbar c}$ and putting
the latter equal to one.
}

We still have to choose a suitable gauge for the radial coordinate $R$.
The most natural choice is $R\equiv r$, i.e.\ Schwarzschild
(S) coordinates.
However, this choice is singular at a stationary point of $r(R)$
(`equator').
Such singularities are avoided using the gauge
$B\equiv r^{-2}$ for $B>0$ resp.\ $B\equiv -r^{-2}$ for $B<0$.
We denote this radial coordinate by $\tau$ in order to distinguish it from
the S~coordinate $r$.
With this choice the spatial part of the metric takes the simple form
\begin{equation}
ds^2=r^2(d\tau^2+d\Omega^2)\;,
\end{equation}
suggesting to call them isotropic coordinates.

Using S~coordinates the field equations obtained from
(\ref{Action}) are
\begin{subeqnarray}\label{feq}
(BW')'&=&W({W^2-1\over r^2}+H^2)-2rBW'V_1\;,\\
(r^2BH')'&=&(2W^2+{\beta^2r^2\over2}(H^2-\alpha^2))H-2r^3BH'V_1\;,\\
(rB)'&=&1-2r^2BV_1-2V_2\;,\\
A'&=&2rV_1A\;.
\end{subeqnarray}
The equations obtained with isotropic coordinates
$B\equiv -r^{-2}$ are essentially Eqs.~(9) of \cite{BFM2},
there are however some sign changes due to $B<0$.
\begin{subeqnarray}\label{taueq}
   r'&=&rN\;,\\
   N'&=&(\kappa-N)N-2U^2-V^2\;,\\
  \kappa '&=&-1-\kappa^2+2U^2+{\beta^2r^2\over2}(H^2-\alpha^2)^2+
  2H^2W^2\;,\\
  W'&=&rU\;,\\
  U'&=&-{W(W^2-1)\over r}-rH^2W-(\kappa-N)U\;,\\
  H'&=&V\;,\\
  V'&=&-{\beta^2r^2\over2}(H^2-\alpha^2)H-2W^2H-\kappa V\;,
\end{subeqnarray}
together with the constraint
\begin{equation}\label{kappeq}
2\kappa N=-1+N^2+2U^2+V^2+2V_2\;,
\end{equation}
and the definitions
\begin{equation}
N\equiv{r'\over r}\;,\quad \kappa\equiv{(A^2B)'\over 2A^2B}+N\;,
 \quad U\equiv{W'\over r}\;,\quad{\rm and}\quad V\equiv H'\;.
\end{equation}

\section{Singular points}\label{chapsing}

The field Eqs.~(\ref{feq}) are singular at $r=0$, $r=\infty$ and for
points where $B$ vanishes. Whereas the former singularities are of
geometrical origin (the action of the rotation group degenerates)
the zeros of $B$ turn out to be coordinate singularities, in fact, of
two different kinds:
\begin{enumerate}
\item[1.]
Points where $B=0$, but $A^2B\neq 0$ are stationary points of the
function $r(R)$ and hence choosing $r$ as a coordinate leads to singular
derivatives; our second coordinate choice avoids this problem.
\item[2.]
Points where $B=0$ and $A^2<\infty$ correspond to an horizon;
as is well known a retarded time coordinate $t^*=t-\int{dr\over A|B|}$
avoids this singularity.
\end{enumerate}
As we want to stick to the time coordinate $t$ we have to treat horizons as
singular points. In order to guarantee the finiteness of $A$ we have to
require the regularity conditions
\begin{equation}\label{bc}
  rB'|_h=1-2V_2|_h\;,\qquad
  B'W'|_h={1\over2}
    \left.{\partial V_2\over\partial W}\right|_h\;,\qquad
  r^2B'H'|_h=\left.{\partial V_2\over\partial H}\right|_h\;.
\end{equation}

In general it is not possible to parametrize regular solutions
directly by their initial data at the singular point.
However, as was already discussed in \cite{BFM,BFM1,BFM2} in the present
case this is in fact possible, using a sharpened version
(Prop.~1 of \cite{BFM}) of the standard text-book existence theorems
\cite{Codd}.

\noindent
{\sl Proposition}:\\
Consider a system of
first order differential Eqs.\ for $m+n$ functions $y=(u,v)$
\begin{subeqnarray}\label{lin2}
s{du_i\over ds}&=&s f_i(s,y)\;,\qquad i=1,\ldots,m\;,\\
s{dv_i\over ds}&=&-\lambda_i v_i + s g_i(s,y)\;,\qquad i=1,\ldots,n\;,
\end{subeqnarray}
with constants $\lambda_i>0$ and let $\cal C$ be an open subset of ${\bf R}^m$
such that $f_i$, $g_i$ are analytic in a neighbourhood of $s=0$,
$y=(c,0)$ for all $c\in\cal C$. There exists an $m$-parameter family of
local solutions $y_c(s)$ analytic in $c$ and $s$ for $c\in\cal C$,
$|s|<s_0(c)$ such that $y_c(0)=(c,0)$.

In order to meet the requirements of this Proposition we may introduce
the coordinate $s\equiv r-r_h$ and put \cite{BFM1}
\begin{subeqnarray}
  u_1&\equiv& r\;,\qquad u_2\equiv W\;,\qquad
  u_3\equiv H\;,\\
  v_1&\equiv&{B\over s}-{1\over u_1}\biggl(1-2V_2\biggr)\;,\\
  v_2&\equiv&{BW'\over s}-
              u_2\biggl({u_2^2-1\over u_1^2}+u_3^2\biggr)\;,\\
  v_3&\equiv&{r^2BH'\over s}-u_3\biggl(2u_2^2
           +{\beta^2\over2}u_1^2(u_3^2-1)\biggr)\;.
\end{subeqnarray}
There remain two parameters --
$W_h$ and $H_h$ -- to describe solutions with a regular
horizon at $r_h$. For an event horizon $r$ has to increase
as one deviates from the horizon in the direction where $B$ is positive.
This requires the inequality
\begin{equation}\label{pos}
rB'|_h=1-2V_2|_h>0\;,
\end{equation}
which for the case without Higgs field reduces to $r_h>|W_h^2-1|$.  For
$V_2|_h=1/2$ we get $B'|_h=0$, which looks like the condition
for a degenerate horizon.  As discussed in \cite{BFM2} the latter is,
however, only obtained for
$1-2V_2|_h=\partial V_2/\partial W|_h=\partial V_2/\partial H|_h=0$; without
Higgs field this is satisfied only for $r_h=1$, $W_h=0$ yielding the
extremal Reissner-Nordstr{\o}m solution.  Initial data with $V_2|_h=1/2$ but
$\partial V_2/\partial W|_h\ne0$ and\slash or
$\partial V_2/\partial H|_h\ne0$ turn out to describe a non-degenerate
horizon, which is simultaneously a maximum of $r$ and therefore has a
singularity in S~coordinates.
In order to avoid this singular behavior one may use isotropic
coordinates, in which the boundary conditions at the horizon (assumed to
be at $\tau=0$) read \cite{BFM2}
\begin{subeqnarray}\label{bcbh}
  r(\tau)&=&r_h\left(1+N_1\left({\tau^2\over2}+O(\tau^4)\right)
    -\left(W_1^2+{1\over2}H_1^2\right){\tau^4\over4}\right)+O(\tau^6)\;,
  \qquad\\
  N(\tau)&=&N_1(\tau+O(\tau^3))
    -\left(W_1^2+{1\over2}H_1^2\right)\tau^3+O(\tau^5)\;,\\
  W(\tau)&=&W_h+r_h W_1{\tau^2\over2}+O(\tau^4)\;,\\
  H(\tau)&=&H_h+H_1{\tau^2\over2}+O(\tau^4)\;,
\end{subeqnarray}
where
\begin{equation}\label{w1def}
  N_1=-{1\over2}+\left.V_2\right|_h\;,\qquad
  W_1=-{r_h\over4}\left.\partial V_2\over\partial W\right|_h\;,\qquad
  H_1=-{1\over2}\left.\partial V_2\over\partial H\right|_h\;,
\end{equation}
The function $\kappa(\tau)$ has a simple pole at the horizon, but
$\kappa(\tau)-1/\tau$ is regular.
$B\equiv-N^2$, $W$, and $H$ are analytic in $\tau^2$ and $\tau^2$ is
analytic in the S~coordinate $r$ as long as $V_2|_h\ne1/2$ (i.e.,
$N_1\ne0$). For the special case $V_2|_h=1/2$ we obtain
\begin{subeqnarray}\label{sbc}
  B&=&-8(W_1^2+H_1^2/2)^{1\over2}(1-r/r_h)^{3\over2}+O((1-r/r_h)^2)\;,\\
  W&=&W_h+r_hW_1\left(1-r/r_h\over W_1^2+H_1^2/2\right)^{1\over2}
    +O(1-r/r_h)\;,\\
  H&=&H_h+H_1\left(1-r/r_h\over W_1^2+H_1^2/2\right)^{1\over2}
    +O(1-r/r_h)\;.
\end{subeqnarray}

In the following we will use the terms event resp.\ Cauchy horizon for any
horizon with $B'|_h>0$ resp.\ ${}<0$, although the original meaning of these
terms applies only to asymptotically flat solutions.

Next we turn to the singular behaviour at $r=0$, which is of particular
relevance for the internal structure of black hole solutions.
We have to distiguish two cases, $B>0$ and $B<0$.
\begin{enumerate}
\item[1.]{\bf $B>0$}:\qquad
For black holes this case is only possible, if there is a second, inner
horizon.
The generic behaviour of the solutions in isotropic coordinates
without the Higgs field is
described by Prop.~13 of \cite{BFM}. Translating to S~coordinates
and taking into account the Higgs field we get
\begin{subeqnarray}\label{bcrn}
  W(r)&=&W_0+{W_0\over 2(1-W_0^2)}r^2+W_3r^3+O(r^4)\;,\\
  H(r)&=&H_0+H_1r+O(r^2)\;,\\
  B(r)&=&{(W_0^2-1)^2\over r^2}-{2M_0\over r}+O(1)\;,
\end{subeqnarray}
which is a 5-parameter, i.e.\ generic, family of solutions.
In the case with vanishing Higgs field Prop.~13 of \cite{BFM}
implies the analyticity of $W(r)$ and $r^2B(r)$, i.e.\ the
expressions above are in fact the beginning of a convergent Taylor
series.
More generally, in the case with a Higgs field,
introducing the variables
\begin{subeqnarray}
  u_1&\equiv& W\;,\qquad u_2\equiv (rBW'+W(W^2-1))/r\;,\\
  u_3&\equiv& \biggl({(W^2-1)^2\over r^2B}-1\biggr)/r\;,\\
  u_4&\equiv& H\;,\qquad u_5\equiv H'\;,
\end{subeqnarray}
it is easy so check that
\begin{equation}
ru_i'=O(r)\;, \qquad i=1\dots 5\;,
\end{equation}
and thus analyticity follows from the Proposition with
${\cal C}=\{W_0^2\neq 1\}$.

According to the asymptotics of $B(r)$ we may call the singular
behaviour to be of RN-type.
The special case $W_0^2=1, M_0<0$ leads to S-type behavior with a
naked singularity.
On the other hand $W_0^2=1, H_0=0, M_0=0$ gives regular solutions.

\item[2.] {\bf $B<0$}:\qquad
This case is more involved, with two disjoint families of singular
solutions.

\item[2.1]
There is a 3-parameter family of solutions with a S-type singularity,
characterized by the asymptotics
\begin{subeqnarray}\label{bcss}
  W(r)&=&1+W_2r^2+O(r^3)\;,\\
  H(r)&=&H_0+O(r)\;,\\
  B(r)&=&-{2M_0\over r}+O(1)\;.
\end{subeqnarray}
Introducing the variables
\begin{subeqnarray}
  u_1&\equiv& BW'\;,\qquad u_2\equiv{1\over rB}\;,\qquad
  u_3\equiv H\;,\\
  v_1&\equiv&{(1-W^2)\over r^2}+{W'\over 2r}\;,\\
  v_2&\equiv& rBH'-2H\;,
\end{subeqnarray}
we obtain
\begin{subeqnarray}
  ru_i'&=&O(r)\;, \qquad{\rm for}~i=1\dots 3\;,\\
  rv_1'&=&-2v_1+O(r)\;,\qquad rv_2'=-v_2+O(r)\;.
\end{subeqnarray}
With ${\cal C}=\{M_0>0\}$ the analyticity follows again from the
Proposition.

Obviously the condition $B<0$ ($M_0>0$) prevents the existence of
regular solutions in this case.

\item[2.2 ]
There is an additional 2-parameter family of solutions with a pseudo-RN
singularity (pseudo because $B<0$).
\begin{subeqnarray}\label{bcqrn}
  W(r)&=&W_0\pm r+O(r^2)\;,\\
  H(r)&=&H_0+O(r^2)\;,\\
  B(r)&=&-{(W_0^2-1)^2\over r^2}\pm{4W_0(1-W_0^2)\over r}+O(1)\;.
\end{subeqnarray}
Introducing the variables
\begin{subeqnarray}
  u_1&\equiv& W\;,\qquad   u_2\equiv H\;,\\
  v_1&\equiv& \biggl[1\mp \biggl(W'-{W(W^2-1)\over rB}\biggr)\bigg]/r\;,
  \quad v_3\equiv H'\;,\\
  v_2&\equiv& \bigl({(W^2-1)^2\over r^2B}+1\bigr)/r\;,
\end{subeqnarray}
we get
\begin{subeqnarray}
  ru_i'&=&O(r) \quad{\rm for}~i=1,2\;\\
  rv_1'&=&-2v_1-4v_2+O(r)\;,
  \quad rv_2'=v_1-v_2+O(r)\;,\\
  rv_3'&=&-v_3+O(r)\;.
\end{subeqnarray}
With ${\cal C}=\{W_0^2\neq 1\}$ the Proposition implies
again analyticity.
The negative eigenvalues of the linearized equations are
$\lambda_{1,2}=-1/2(3\pm i\sqrt 15)$ and $\lambda_3=-1$.
This is a repulsive focal point which will turn out to be important for the
cyclic behaviour in the EYM case.
\end{enumerate}
In the case $B<0$ we obtained no singular class that has enough parameters
(three for EYM and five for EYMH) to describe the generic behaviour. Since,
also the appearance of a second, inner horizon is a
non-generic phenomenon, one may wonder, what the generic behaviour inside
the horizon near $r=0$ looks like. This question actually is at the origin
of our interest in this problem. Some insight was obtained by a detailed
numerical study, whose results will be presented in the next chapter
(compare also \cite{DGZ}).

Finally we would like to address the question of geodesic incompleteness
inside the horizon expected in view of general singularity theorems \cite{HE}.
We will show that for solutions without a Cauchy horizon (i.e., $B<0$ for
$r<r_h$) the radial time- and lightlike geodesics are incomplete.
Their equations are
\begin{equation}
  g_{tt}\biggl({dt\over d\tau}\biggr)^2
    -g_{rr}\biggl({dr\over d\tau}\biggr)^2=\epsilon \quad
{\rm with}\quad \epsilon=1\;{\rm or}\;0 \;.
\end{equation}
{}From staticity we get the constancy of $g_{tt}dt/d\tau=E$ and thus
\begin{equation}
 \biggl ({dr\over d\tau}\biggr)^2={E^2\over A^2}-\epsilon B \;.
\end{equation}
{}From Eq.~(\ref{feq}d) we see that $A$ is a monotone function of $r$
implying
\begin{equation}
  \tau_h\le\int_0^{r_h}{A\over E}dr\le{A(r_h)\over E}r_h<\infty \;,
\end{equation}
where $\tau_h$ is the difference in proper time resp.\ affine parameter
between the horizon and the origin.

\section{Numerical results}\label{chapnum}

In order to investigate the generic behaviour of non-abelian black
holes inside the event horizon, we integrate the field
Eqs.~(\ref{feq}) resp.\ (\ref{taueq}) from the horizon assuming
$B<0$, ignoring the constraints on the initial data at the
horizon required for asymptotic flatness.

\begin{figure}[t]
\hbox to\hsize{\hss
   \epsfig{file=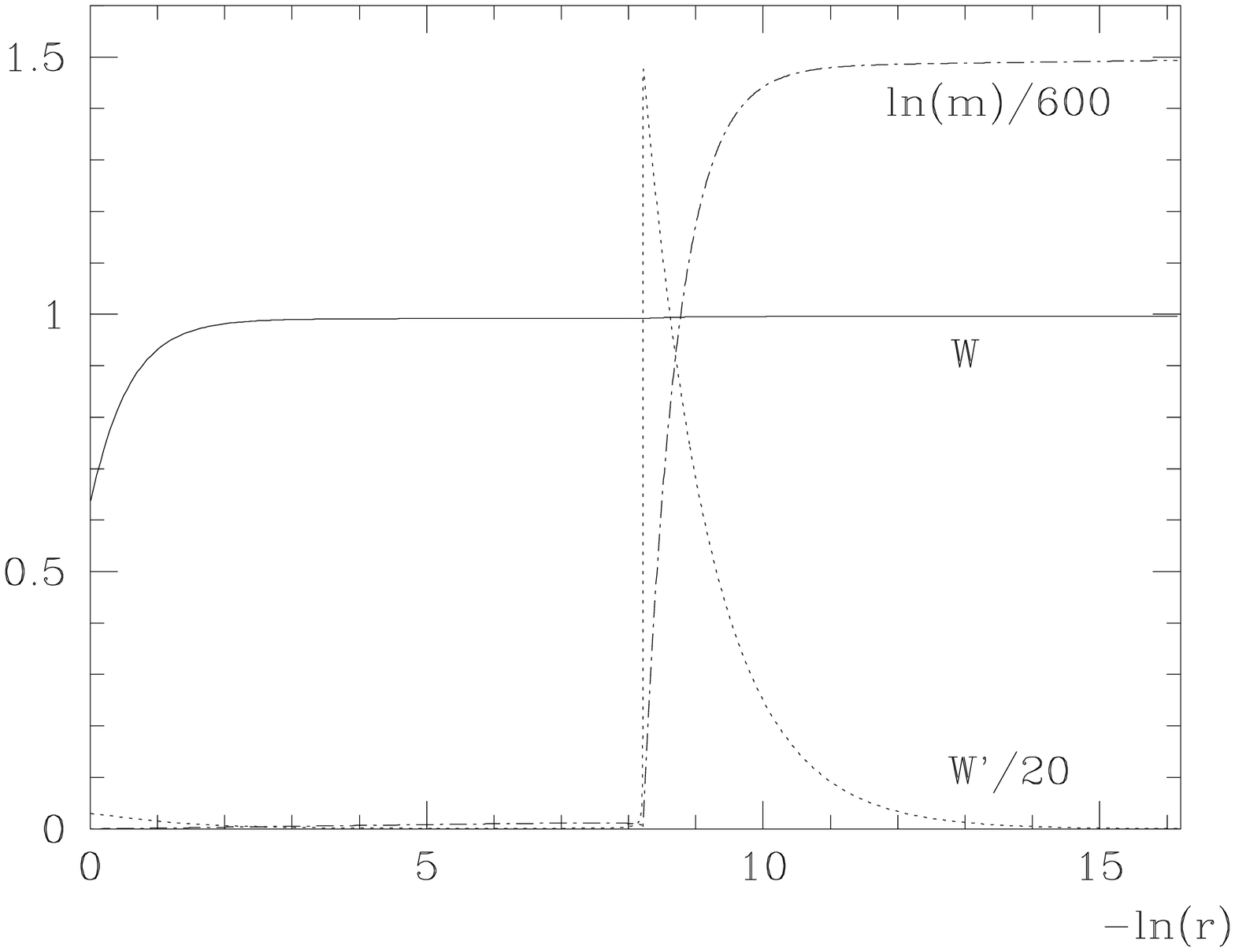,width=0.6\hsize,%
   bbllx=1.3cm,bblly=4.9cm,bburx=20.5cm,bbury=18cm}%
  \hss}
\labelcaption{expl1}{
First inflationary cycle of the fundamental black hole of EYM theory
with $r_h=1$ and $W_h=0.6322\,.$}
\end{figure}

As mentioned before, the
horizon is a singular point of the equations. Consequently one has to
desingularize the equations in order to be able to start the integration
right there. How this can be done, was described in the previous chapter
resp.\ in \cite{BFM2,BFM1}.

If one ignores the behaviour of the solutions outside the horizon, one
may relax the condition (\ref{pos}) ensuring a decreasing
resp.\ increasing $r$ inside resp.\ outside the horizon.
At least part of the solutions violating (\ref{pos}) possess an equator,
i.e.\ a maximum of $r$. For those the use of S~coordinates is excluded.

\begin{figure}
\hbox to\hsize{
  \epsfig{file=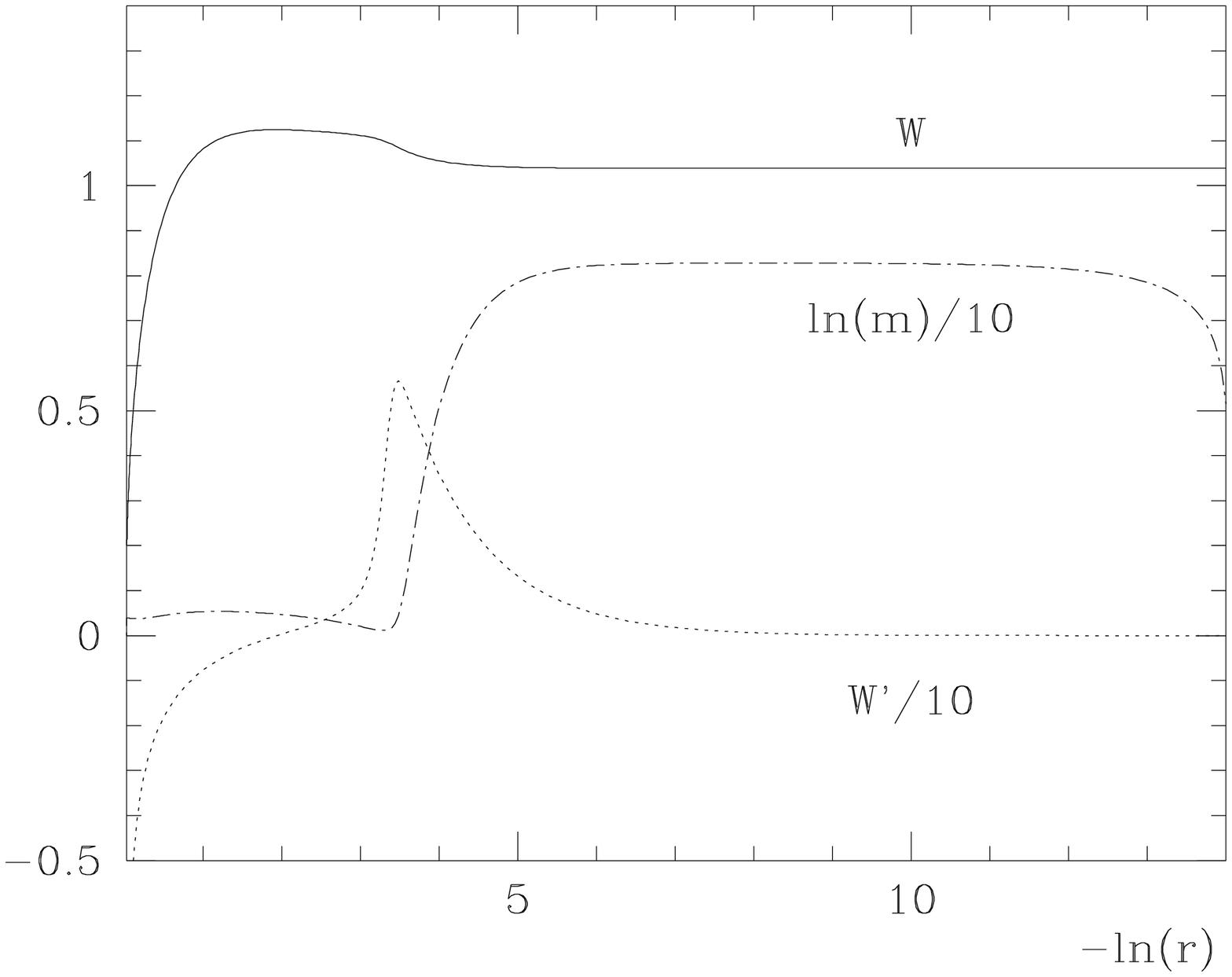,width=0.48\hsize,%
      bbllx=2.3cm,bblly=5.9cm,bburx=20.0cm,bbury=20.0cm}\hss
  \epsfig{file=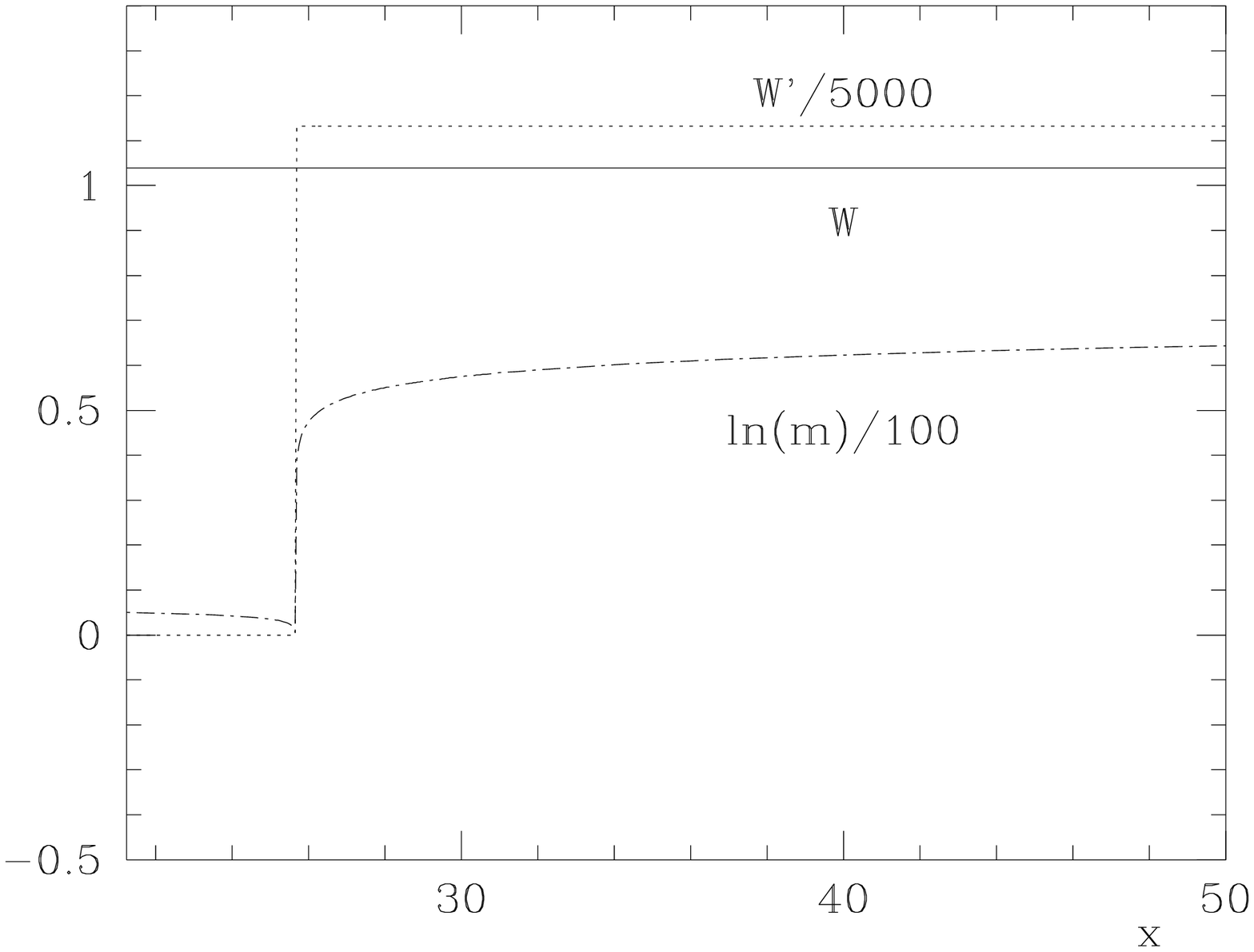,width=0.48\hsize,%
      bbllx=1.8cm,bblly=5.5cm,bburx=20.0cm,bbury=20.0cm}
  }
\labelcaption{expl2}{
First two cycles of the solution with  $r_h$=0.97 and $W_h=0.2$.  For
the second cycle a suitably stretched coordinate $x$ is used.}
\end{figure}

As one performs the numerical integration one quickly runs into serious
problems due to the occurence of a quasi-singularity, initiated
by a sudden steep raise of $W'$ and subsequent exponential growth
of $B$ resp.\ $m$ (compare Figs.~\ref{expl1}, \ref{expl2}, and~\ref{explh}
for some examples).
This inflationary behaviour of the mass function
is similar to the one observed for perturbations of the RN solution
at the Cauchy horizon \cite{IP}. While this exponential growth continues
indefinitely for the EYMH system, it comes to a stop without the Higgs
field. The mass function reaches a plateau and stays constant for a
while until it starts to decrease again. When $B$ has become small
enough, i.e.\ the solution comes close to an inner horizon, the same
inflationary process repeats itself. Generically this second
``explosion'' is so violent (we will give estimates on the increase
of $m$ in chapter~\ref{chapqual}) that the numerical integration
procedure breaks down.

Besides these generic solutions there are certain families of special
solutions obtained through suitable fine-tuning of the initial data at
the horizon.  There are two classes of such special solutions.  The
first class are black holes with a second, inner horizon, the second are
solutions with one of the singular behaviours at $r=0$ for $B<0$
described in chapter~\ref{chapsing}.  The numerical construction of such
solutions is complicated by the fact that both boundary points are
singular points of the equations.  The strategies employed to solve such
problems are well described in our paper on gravitating monopoles
\cite{BFM2}.  Actually, in order to control the numerical uncertainties
we used two different methods, which may be called ``matching'' and
``shooting and aiming''.  For matching we integrate independently from
both boundary points with regular initial data, tuning these data at
both ends until the two branches of the solution match.  For shooting
and aiming we integrate only from one end and try to suppress the
singular part of the solution at the other end by suitably tuning the
initial data at the starting point.

\begin{figure}
\hbox to\hsize{
  \epsfig{file=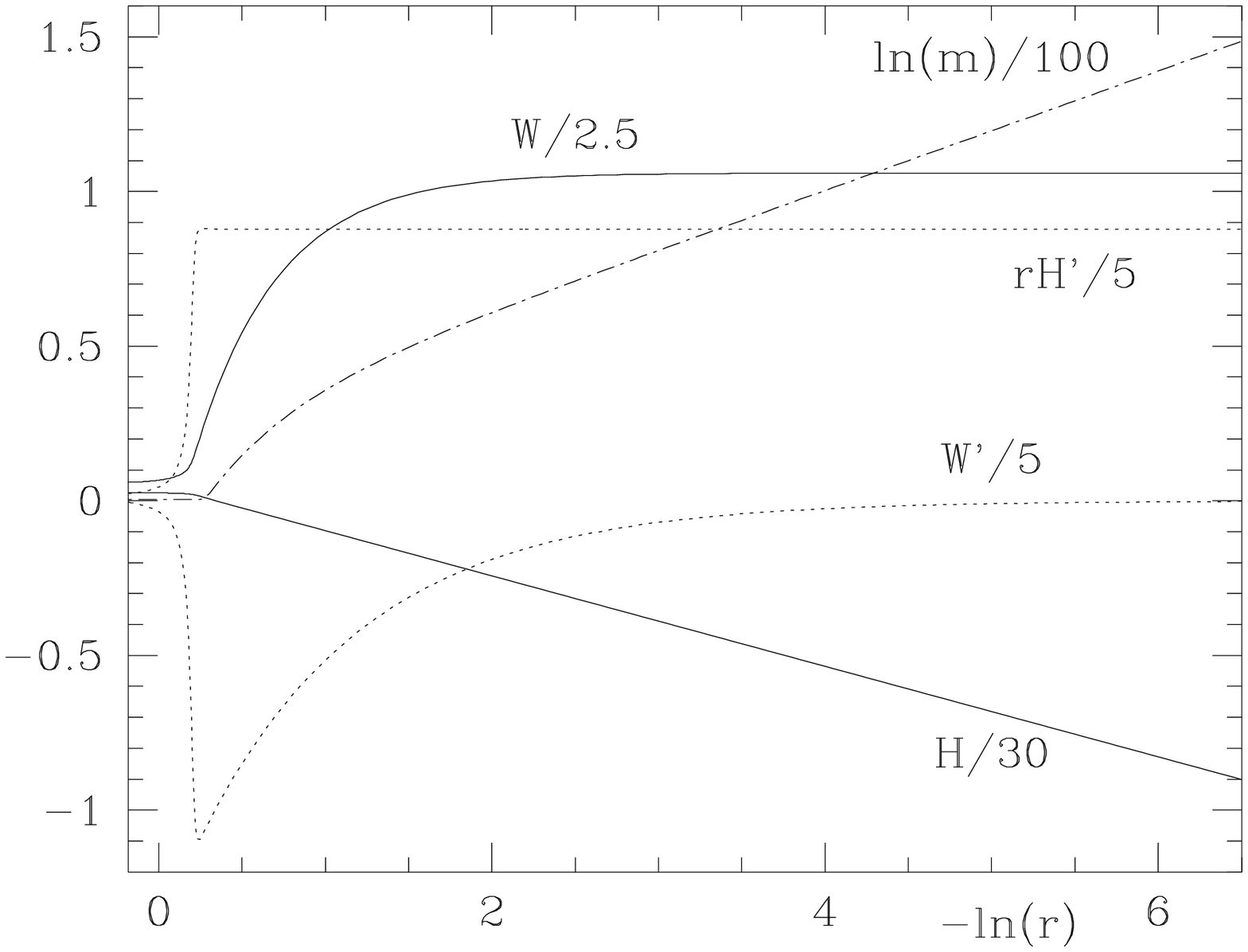,width=0.48\hsize,%
      bbllx=2.3cm,bblly=5.9cm,bburx=20.0cm,bbury=20.0cm}\hss
  \epsfig{file=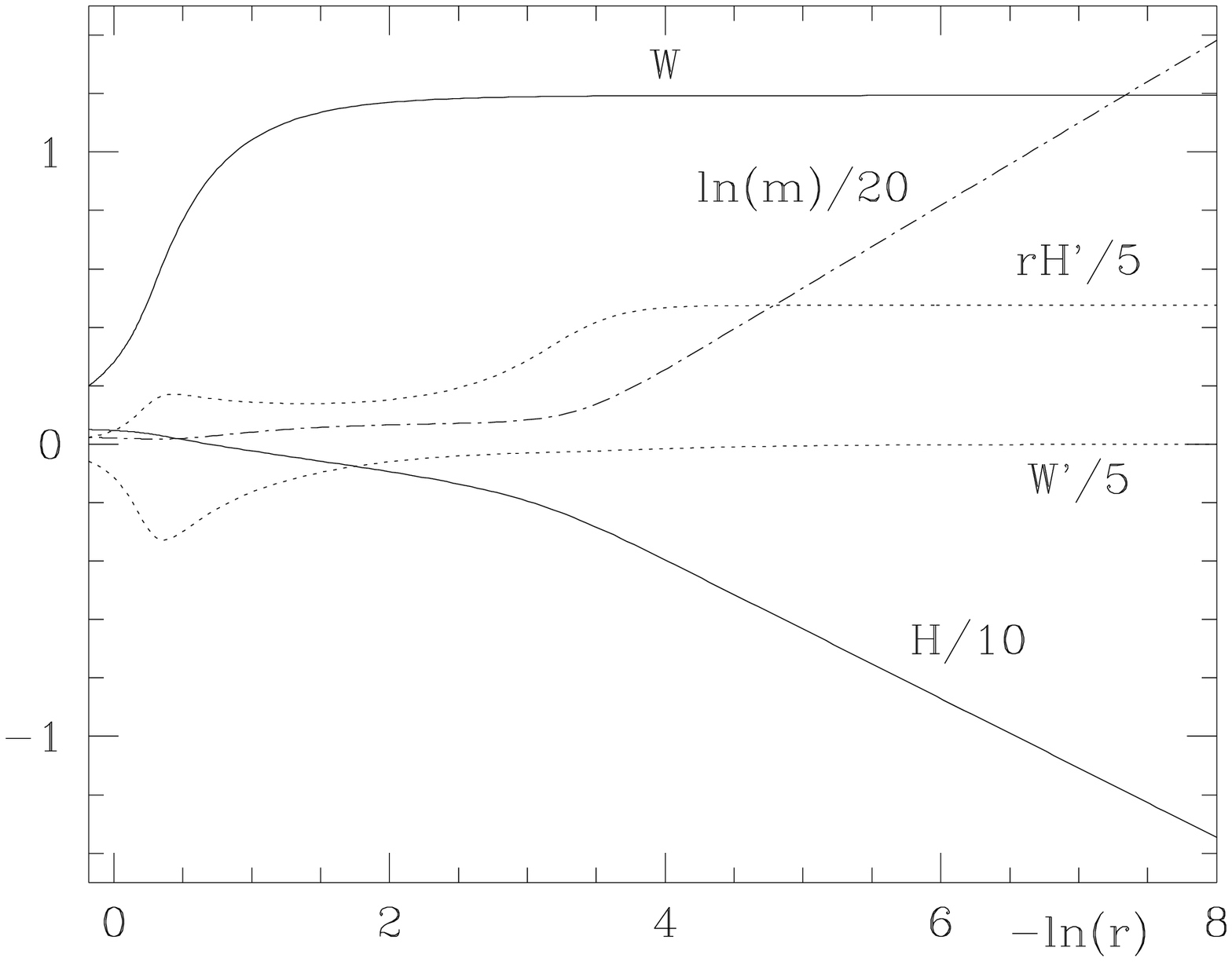,width=0.48\hsize,%
      bbllx=1.8cm,bblly=5.5cm,bburx=20.0cm,bbury=20.0cm}
  }
\labelcaption{explh}{
Inflationary solutions with Higgs fields with $\alpha=0.2$, $\beta=0$,
$r_h=1.2$ and $W_h=0.15$, $H_h=0.8$ resp.\ $W_h=0.2$, $H_h=0.5\,.$}
\end{figure}
As already said, the first class of special solutions
consists of black holes with a second, inner horizon; let us call them
non-abelian RN-type (NARN) solutions.
As was explained in chapter~\ref{chapsing} a regular horizon
requires the simultaneous vanishing of $B$, $BW'$ and $BH'$.
We have determined two such 1-parameter families for the EYM system,
shown in Fig.~\ref{curves},
whose significance will be explained in chapter~\ref{chapqual}.
As may be inferred from Fig.~\ref{curves}, the (dotted) curve~2
corresponding to one such family intersects all (solid) curves
describing asymptotically flat solutions except the one for $n=1$.
In contrast to what is claimed by Donets et al.\
\cite{DGZ} our curve continues straight through the parabola
$r_h=1-W_h^2$ and runs all the way to $r_h=W_h=0$.
As already mentioned the
branch to the left of the parabola cannot be obtained using
S~coordinates.
The corresponding curve of Donets et al.\ makes a
suspiciously sharp turn very close to the parabola and
runs to the point $r_h=1, W_h=0$. We are convinced that the latter
piece of their curve is an artefact of unreliable numerics caused by the use
of S~coordinates becoming singular at the parabola
(compare the discussion in chapter~\ref{chapsing}).

As already stressed $B'$ has to be positive at an event horizon.
This condition is violated for
values of $r_h<r_p$, where $r_p$ denotes the value of $r_h$ where the
NARN curve~2 intersects the parabola. However it turns out
to be fulfilled for the second horizon if $r_h<r_m$ with
$r_m\approx 0.112$ on the NARN curve. Thus this curve may be divided
into three pieces according to $0<r_h<r_m$, $r_m<r_h<r_p$ and $r_p<r_h$.
On the first interval the two horizons have exchanged their roles, whereas
on the second interval both of them are Cauchy horizons with a maximum of
$r$ (equator) in between. The only difference between the first and the
third interval is that $W_h>1$ resp.\ ${}<1$ on the event horizon.

\begin{figure}
\hbox to\hsize{\hss
   \epsfig{file=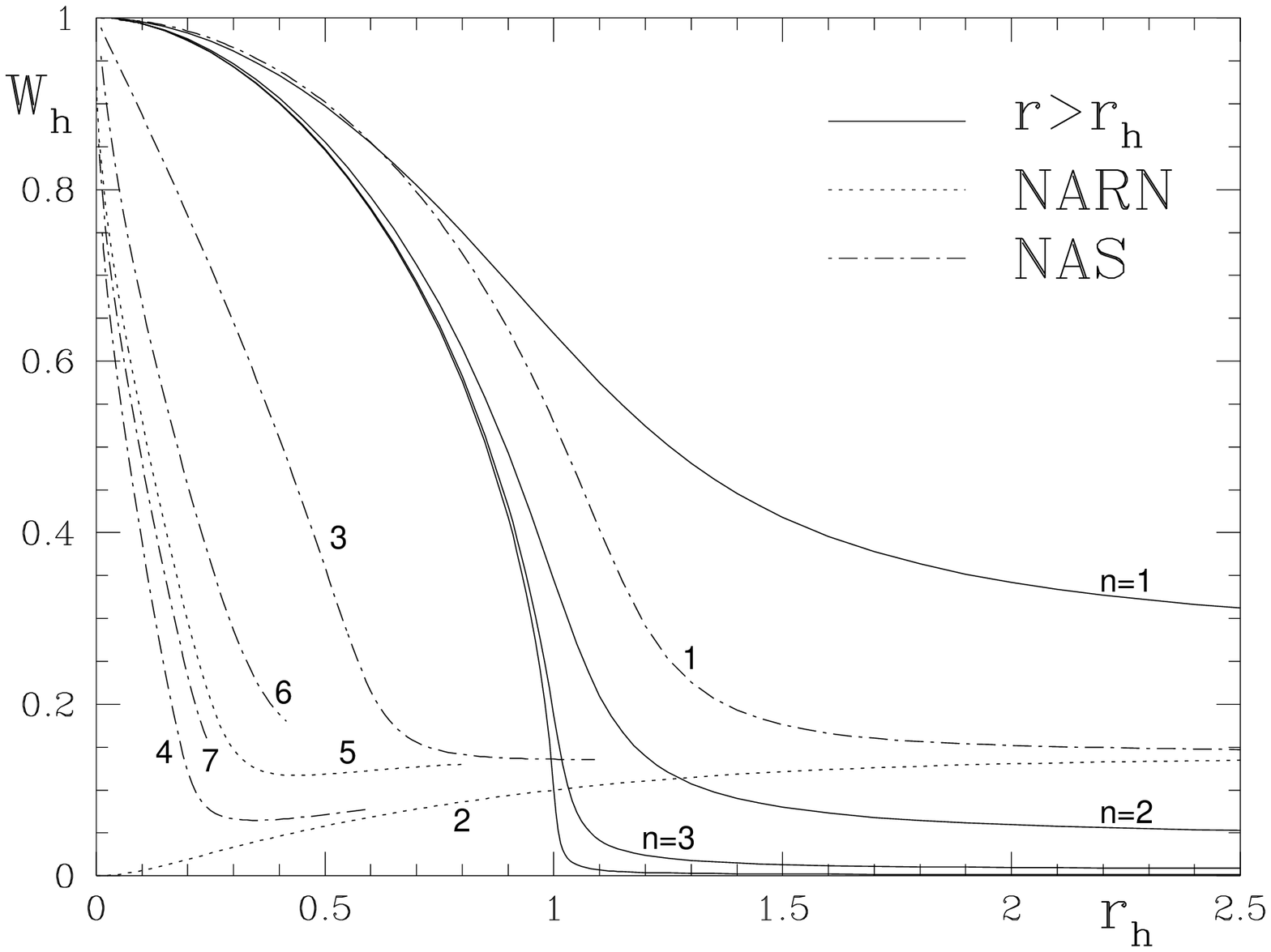,width=\hsize,%
   bbllx=1.3cm,bblly=6.1cm,bburx=20.0cm,bbury=19.8cm}%
  }
\labelcaption{curves}{
Initial data for special solutions. The solid curves represent
asymptotically flat solutions with $n$ zeros of $W$. The other curves
represent various NARN and NAS families.}
\end{figure}

Our second NARN family (curve~5 of Fig.~\ref{curves}) stays
completely to the left of the parabola and ends at $r_h\approx0.9$ close
to the curve~3, whose significance will be explained below.

\begin{figure}
\hbox to\hsize{
  \epsfig{file=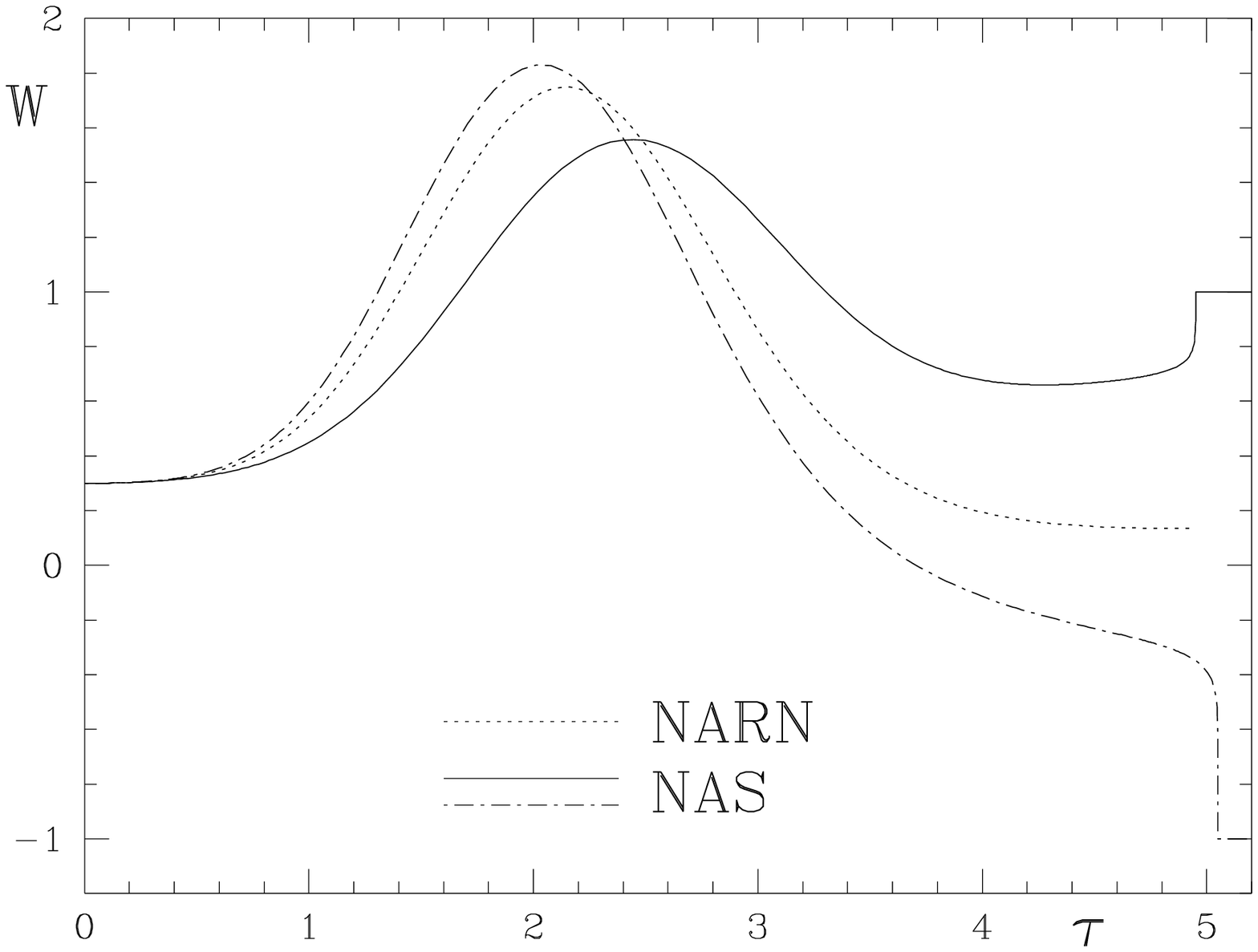,width=0.48\hsize,%
      bbllx=1.8cm,bblly=6.0cm,bburx=19.5cm,bbury=20.0cm}\hss
  \epsfig{file=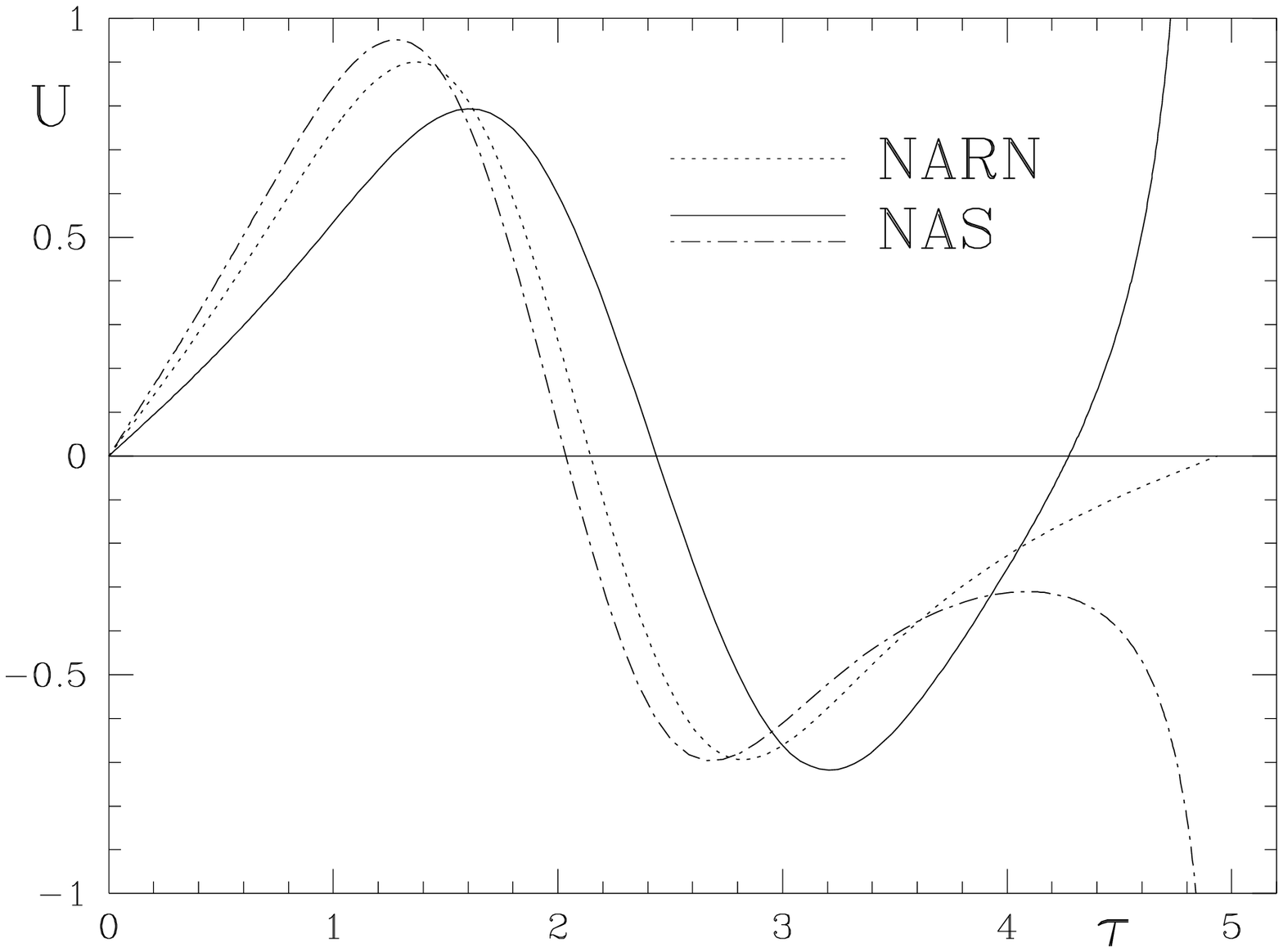,width=0.48\hsize,%
      bbllx=1.8cm,bblly=6.0cm,bburx=19.5cm,bbury=20.0cm}
  }
\labelcaption{NAS}{
A NARN solution and two accompanying NAS solutions with $W_h=0.3$ and
$r_h=0.198728$ (on curve~5 of Fig.~\ref{curves}), $r_h=0.290837$ (on curve~6
with $W\to+1$), and $r_h=0.170525$ (on curve~7 with $W\to-1$).}
\end{figure}

The second class are solutions without a second horizon
(i.e.\ $B$ stays negative) approaching the center $r=0$ with one of the
two singular behaviours described in chapter~\ref{chapsing},
i.e.\ those with a S-type singularity resp.\ with a pseudo-RN-type
singularity; let us denote them NAS resp.\ NAPRN solutions.
We have determined several NAS families
represented by the dashed-dotted curves of Fig.~\ref{curves}.
The curve~1 staying to the right of the parabola coincides with the
corresponding one of Donets et al., whereas the others, staying
essentially to the left of the parabola are new. As will be explained in
chapter~\ref{chapqual}, the two NAS curves~6 and~7 accompanying the (dotted)
NARN curve~5 are expected to merge with the NAS curve~3 close to
$r_h=0.9$. Some of the NAS curves (e.g., 3 and~4) are expected to extend
indefinitely to the right, but numerical difficulties (too violent
``explosions'') prevented us from continuing them further to larger values
of $r_h$. They will intersect the (solid) curves for asymptotically flat
solutions with $n=2,3,\ldots$ zeros of $W$ and therefore yield additional
asymptotically flat NAS black holes beside those found by Donets et
al.\ \cite{DGZ}, contradicting their uniqueness claim.

Finally there are the NAPRN solutions, which constitute a discrete set
according to the number of available free parameters at $r=0$.
We found several such solutions with $W'(0)=-1$ (compare Tab.~\ref{NAPRN}).
Only one of them has no maximum of $r$ and was also found by Donets et al.

\begin{table}[t]
\labelcaption{NAPRN}{
Initial data for several NAPRN solutions with $W'(0)=-1\,.$}
\begin{center}\leavevmode
\input w0rhwh.out
\end{center}
\end{table}

\section{Qualitative Discussion}\label{chapqual}

We shall now give a qualitative picture of the solutions and try to
explain our numerical results. Since the generic behaviour of the
solutions is rather different in the cases with and without Higgs field,
we shall treat the two cases separately. Let us first concentrate
on the case without Higgs field.
For simplicity we introduce the notation
$\bu\equiv BW'$ and $\bb\equiv rB$ and use $\sigma\equiv -\ln(r)$
as a radial coordinate. Observe that $\bb\approx -2m$ for small $r$.
With these variables the field Eqs.~(\ref{feq})
become (a dot denoting $d/d\sigma$)

\begin{subeqnarray}\label{bareq}
\dot{W}&=&-r^2{\bu\over\bb}\;,\\
\dot{\bu}&=&-W{W^2-1\over r}+
2r^2{\bu^3\over\bb^2}\;,\\
\dot{\bb}&=&r\biggl({(1-W^2)^2\over r^2}-1\biggr)+2r^2{\bu^2\over\bb}\;.
\end{subeqnarray}
Close to the horizon the first term in the equation for $\bb$
dominates (since $\bu$ vanishes at $r=r_h$) and thus $\bb$
becomes negative. Provided $W^2$ does not tend to $1$,
this term will, however, change sign as $r$ decreases and $\bb$ will turn back to
zero.
Assuming further that $\bu$ does not tend to zero simultaneously,
the second term in the equation for $\bu$ will grow very rapidly
as $\bb$ approaches zero, leading to a rapid increase of $\bu$.
This in turn induces a rapid growth of $\bb$ (compare Fig.~\ref{expl1}).
Once the second terms in Eqs.~(\ref{bareq}b,c)
dominate one gets
${(\bu/\bb)}\dot{}\approx 0$ and thus $\bu/\bb=W'/r$ tends to a constant
$c$.
As long as $(rc)^2$ is sizable $\bu$ and $\bb$ increase exponentially,
giving rise to the phenomenon of mass inflation.
Eventually this growth comes to a stop when $(rc)^2$ has
become small enough. Then $\bu$ and $\bb$ stay constant until
the first terms in Eqs.~(\ref{bareq}b,c) become sizable again.
As before $\bb$ tends to zero inducing another ``explosion'' resp.\
cycle of mass inflation (compare Fig.~\ref{expl2}).

In the discussion above we made two provisions -- that
$W^2$ stays away from $1$ and that $\bu$ does not tend zo zero
simultaneously with $\bb$. If the first condition is violated, i.e.\
$W^2\to 1$ we get a NAS solution. If on the other hand $\bu$ and $\bb$
develop a common zero we get a NARN solution, i.e.\ a solution
with a second horizon. Both these phenomena can occur after any finite
number of cycles, giving rise to several NAS resp.\ NARN curves as in
Fig.~\ref{curves}.
Generically $W$ changes very little during an inflationary cycle,
with the exception of solutions that come very close to a second
horizon, i.e.\ close to a NARN solution. In this case $W$ may change by
any amount, depending on how small $\bu$ becomes at the start of the
explosion. By suitably fine-tuning the initial data at the horizon one
can then obtain new NAS solutions with $W\to\pm1$ or a new NARN solution. In
this way each NARN solution is the `parent' of two NAS and one NARN
solution. Fig.~\ref{curves} shows two such generations: the NARN solutions
labelled~2 have the NAS children~3 and~4 and the NARN child~5; the curves
labelled~6 and~7 are the NAS children of~5 (see Fig.~\ref{NAS}).
Whenever the value of $W$ at the second horizon of a NARN solution approaches
$\pm 1$ this NARN curve and its NAS children merge with the corresponding
sibling NAS curve having one cycle less. This hierarchy of special solutions
gives rise to a kind of chaotic structure in this region of `phase space''.

After this qualitative discussion we would like to present a simplified
quantitative discussion of the solutions, following essentially one
complete cycle. This will also provide us with a discrete map of the
variables $W,\bu$ and $\bb$ from one plateau to the next.
Let us denote their initial data at some plateau by
$W_0,\bu_0$ and $\bb_0$.
Since the plateau is characterized by the (effective) vanishing
of the second terms in Eqs.~(\ref{bareq}b,c) involving the factor $r^2$
and the constancy of $W$, we can easily integrate them to
\begin{equation}\label{approx}
\bu=\bu_0-{W_0(W_0^2-1)\over r}\;,
\quad{\rm and}\quad
\bb=\bb_0+{(W_0^2-1)^2\over r}\;,
\end{equation}
ignoring the term $-r$ in the $\bb$ equation.
The beginning of the subsequent explosion is characterized by
$\bb\approx 0$, i.e.\ $r\approx r_0=-(W_0^2-1)^2/\bb_0$.
Depending on the value of $W_0$ the value of $\bu$ at this point is
essentially given by $\bu_0$ or $-W_0(W_0^2-1)/r_0$. Whatever it is,
let us denote this value again by $\bu_0$.
For the description of the solution through the explosion it is
sufficient to solve the simplified system
\begin{equation}\label{leadeq}
\dot W=-r^2{\bu\over\bb}\;,\quad
\dot\bu=2r^2{\bu^3\over\bb^2}\;,\quad
\dot\bb=2r^2{\bu^2\over\bb}\;,
\end{equation}
implying $\bu/\bb=c$ with some constant $c$.
Plugging $c$ back into Eqs.~(\ref{leadeq})
we can integrate them to get
\begin{subeqnarray}\label{expsol}
\bu&=&\bu_0\exp\biggl(c^2(r_0^2-r^2)\biggr)\;,\\
\bb&=&{\bu_0\over c}\exp\biggl(c^2(r_0^2-r^2)\biggr)\;,\\
W&=&W_0+{c\over2}(r^2-r_0^2)\;.
\end{subeqnarray}
We still have to determine $c$ joining this solution to the one before
the explosion. Equating the derivatives of $\bb$ at $r=r_0$ using the
expressions in Eqs.~(\ref{approx}) and (\ref{expsol}) we obtain
$c={(W_0^2-1)^2\over2\bu_0r_0^3}$.

In order to obtain the values of $W_1,\bu_1,\bb_1$ after the explosion
we may savely put $r=0$ in Eqs.~(\ref{expsol}) and obtain
\begin{subeqnarray}\label{map}
\bu_1&=&\bu_0e^{(cr_0)^2}\;,\quad \bb_1={\bu_0\over c}e^{(cr_0)^2}\;,\\
W&=&W_0-{c\over2}r_0^2\;,\quad{\rm with}\quad
   r_0=-{(W_0^2-1)^2\over\bb_0}\;,\quad
c={(W_0^2-1)^2\over2\bu_0r_0^3}\;.
\end{subeqnarray}
It is instructive to illustrate these relations on an example.
We take the fundamental black hole solution with $r_h=1$ and
$W_h=0.6322$ shown in Fig.~\ref{expl1}.
For the first explosion one finds the parameters
$r_0\approx 2.7\cdot 10^{-4}$
and $c\approx 1.1\cdot 10^5$
yielding $cr_0\approx 30$ and thus $\bb_1\sim e^{900}$ and
$W_1-W_0\approx 4\cdot 10^{-3}$.
The subsequent explosion will then take place at the fantastically small
value $r_0\sim e^{-900}\approx 10^{-330}$.

Since the change of $W$ in one inflationary cycle has an extra factor
$r_0$ the function $W$ stays practically constant. If we furthermore concentrate
on cases, where the first term in Eq.~(\ref{bareq}b) can be neglected
we may use the simplified system
\begin{equation}\label{simpeq}
\dot W=0\;,\quad
\dot\bu=2r^2{\bu^3\over\bb^2}\;,\quad
\dot\bb={(1-W^2)^2\over r}+2r^2{\bu^2\over\bb}\;,
\end{equation}
also discussed by Donets et al.~\cite{DGZ}.
Introducing the variables $x\equiv r\bu/\bb=W'$ and
$y\equiv -(1-W^2)^2/r\bb$ one obtains the autonomous system
\begin{equation}\label{autoeq}
\dot W=0\;,\quad
\dot x=(y-1)x\;,\quad
\dot y=y(y+1-2x^2)\;.
\end{equation}
Since the first of these equations may be ignored, we can concentrate
on the $x,y$ part. As usual for 2-dimensional dynamical systems the
global behaviour of the solutions can be analyzed determining its fixed
points. Since the ``large time'' behaviour $\sigma\to\infty$ corresponds to
the limit $r\to 0$ these fixed points are related to the singular solutions
at $r=0$ discussed in chapter~\ref{chapsing}.
There are essentially three different fixed points.
\begin{enumerate}
\item[1.]
For $y<0$ there is the fixed point $x=0$, $y=-1$ giving the RN type
singularity. Its eigenvalues are $-1$ and $-2$, hence it
acts as an attracting center for $\sigma\to\infty$.
\item[2.]
Then there is the point $x=y=0$, a saddle with eigenvalues $\pm1$.
\item[3.]
In addition there are the points $x=\pm1$, $y=1$ with the eigenvalues
$1/2(1\pm i\sqrt15)$, related to the pseudo-RN type singularity.
This fixed point acts as a repulsive focal point, from which the
trajectories spiral outwards. Since solutions of the approximate system
given by the Eqs.~(\ref{autoeq}) cannot cross the coordinate axes,
solutions in the quadrants $y>0$, $x>0$ resp.\ $x<0$ stay there
performing larger and larger turns around the focal point coming closer
and closer to the saddle point $x=y=0$ without ever meeting it.
As observed by Donets et al.\ this nicely explains the cyclic
inflationary behaviour of the solutions in the generic case.
\end{enumerate}

Finally we come to the black holes with Higgs field.
Apart from the generic solutions there are the special ones approaching
$r=0$ with a singular behaviour described in chapter~\ref{chapsing}.
On the other hand, the generic behaviour is much simpler than in the
previously discussed situation without Higgs field.
An easy way to understand this difference is to derive the analogue of
the simplified system Eqs.~(\ref{autoeq}). Introducing the additional
variable $z\equiv -\dot{H}$ and ignoring again irrelevant
terms one finds
\begin{subeqnarray}\label{autoheq}
\dot{W}&=&0\;,\quad\dot{H}=-z\;,\\
\dot{x}&=&(y-1)x\;,\quad\dot{z}=yz\;\\
\dot{y}&=&y(y+1-2x^2-z^2)\;.
\end{subeqnarray}
Leaving aside the decoupled equations for $W$ and $H$ one may study the
fixed points of the $(x,y,z)$ system.  For $z=0$ one clearly finds the
previous fixed points of the $(x,y)$ system.  However, for $z\neq0$ the
focal point disappears and the only fixed point for $y\ge0$ is $x=y=0$,
$z=z_0$ with some constant $z_0$.  For $z_0^2<1$ this point is a saddle
with one unstable mode, whereas for $z_0^2>1$ it is a stable attractor.
The latter describes the simple inflationary behaviour described in
chapter~\ref{chapnum} and shown in the left part of Fig.~\ref{explh}.
Solutions approaching a fixed point with $z_0^2<1$ eventually run away
from it again and ultimately tend to one with $z_0^2>1$ as shown in the
right part of Fig.~\ref{explh}.  In both cases the Higgs field grows
logarithmically as $r\to 0$.

Finally let us remark that all these solutions are singular at $r=0$,
since a regular origin requires $B(0)=1$.

\section{Summary}\label{chsum}

We study the behaviour of the static, spherically symmetric non-abelian
black holes inside their event-horizon.  We have chosen a Higgs field in
the adjoint representation of $SU(2)$, but we expect similar results for
Higgs fields in other representations, because for $r\to 0$ the
differences do not seem to be relevant.

In the generic case the solutions inside the horizon show a phenomenon
known as ``mass inflation'' from perturbations of the
Reissner-Nordstr{\o}m black holes at their Cauchy horizon.  Whereas for
the black holes with a Higgs field the mass function grows monotonously
once the inflation has started those without Higgs field run through an
infinite sequence of inflationary cycles, which take the form of ever
more violent ``explosions''.  Besides the generic solutions there are
exceptional ones with a Schwarzschild or Reissner-Nordstr{\o}m type
singularity at the center of symmetry.  These exceptional solutions
accumulate in a way that may be interpreted as a kind of ``chaotic''
behaviour.

Since the typical length scale of these black holes is the Planck length
(at least for values of the gauge coupling $g^2/\hbar c$ of order one)
one may question the physical relevance of these results of the
classical theory.  However, there are claims \cite{ODA} that the
phenomenon of mass inflation connected with the Reissner-Nordstr{\o}m
black hole persists in the quantized theory.  Definitely this subject
requires further study.

Apart from that, our results provide a non-perturbative confirmation of the
mass inflation phenomenon observed for perturbations of the RN black hole.
Furthermore we believe that our results are of some interest in
view of the classical singularity theorems of Penrose and Hawking
\cite{HE}.

\section{Acknowledgments}

G.L. is grateful to Theory Group of the MPI f\"ur Physik for the
invitation and kind hospitality during the visit in Nov.\ 1996, when
this work was begun.  He also wants to thank P.H\'aj\'\i\v{c}ek for
critical remarks and comments.

The work of G.L. was supported in part by the Tomalla Foundation and by
the Swiss National Science Foundation.

\end{document}